\documentclass[aps,prl,twocolumn,superscriptaddress,showpacs,letterpaper]{revtex4}

\usepackage{amsmath}
\usepackage{amsfonts}
\usepackage{amssymb}
\usepackage{bm}
\usepackage{enumerate}
\usepackage[dvips]{color,graphicx}
\usepackage{epsfig}
\usepackage[figuresright]{rotating}

\begin{document}

\title{A precise extraction of the induced polarization in the $^4$He($e,e^\prime \vec p\,$)$^3$H reaction}

\newcommand*{\CNU}{Christopher Newport University, Newport News, Virginia 23606}
\newcommand*{\RUTGERS}{Rutgers, The State University of New Jersey, Piscataway, New Jersey 08854}
\newcommand*{\CSU}{California State University, Los Angeles, Los Angeles, California 90032}
\newcommand*{\DU}{Dalhousie University, Halifax, Nova Scotia, Canada}
\newcommand*{\DUKE}{Duke University, Durham, North Carolina 27708}
\newcommand*{\MIT}{Massachusetts Institute of Technology, Cambridge, Massachusetts 02139}
\newcommand*{\KENT}{Kent State University, Kent, Ohio 44242}
\newcommand*{\WM}{College of William and Mary, Williamsburg, Virginia 23187}
\newcommand*{\ODU}{Old Dominion University, Norfolk, Virginia 23529}
\newcommand*{\ROME}{INFN Rome, Sanit\'a Group and Istituto Superiore di Sanit\'a, I-00161 Rome, Italy}
\newcommand*{\UVA}{University of Virginia, Charlottesville, Virginia 22904}
\newcommand*{\KIPT}{Kharkov Institute of Physics and Technology, Kharkov 310108, Ukraine}
\newcommand*{\NSU}{Norfolk State University, Norfolk, Virginia 23504}
\newcommand*{\SMU}{Saint Mary's University, Halifax, Nova Scotia, Canada}
\newcommand*{\SNU}{Seoul National University, Seoul, Korea}
\newcommand*{\TELAVIV}{Tel Aviv University, Tel Aviv 69978, Israel}
\newcommand*{\UCM}{Universidad Complutense de Madrid, E-28040 Madrid, Spain}
\newcommand*{\ARGONNE}{Argonne National Laboratory, Argonne, Illinois 60439}
\newcommand*{\HU}{Hampton University, Hampton, Virginia 23668}
\newcommand*{\SCAROLINA}{University of South Carolina, Columbia, South Carolina 29208}
\newcommand*{\JLAB}{Thomas Jefferson National Accelerator Facility, Newport News, Virginia 23606}
\newcommand*{\GWU}{The George Washington University, Washington, DC 20052}
\newcommand*{\WEIZ}{Weizmann Institute of Science, Rehovot 76100, Israel}
\newcommand*{\IEM}{Instituto de Estructura de la Materia, CSIC, E-28006 Madrid, Spain}

\author {S.P.~Malace} \affiliation{\SCAROLINA}
\author {M.~Paolone} \affiliation{\SCAROLINA}
\author {S.~Strauch} \affiliation{\SCAROLINA}
\author {I.~Albayrak} \affiliation{\HU}
\author {J.~Arrington} \affiliation{\ARGONNE} 
\author {B.L.~Berman} \affiliation{\GWU}
\author {E.J.~Brash} \affiliation{\CNU} 
\author {B.~Briscoe} \affiliation{\GWU}
\author {A.~Camsonne} \affiliation{\JLAB}
\author {J.-P.~Chen} \affiliation{\JLAB}
\author {M.E.~Christy} \affiliation{\HU}
\author {E.~Chudakov} \affiliation{\JLAB}
\author {E.~Cisbani} \affiliation{\ROME}
\author {B.~Craver} \affiliation{\UVA}
\author {F.~Cusanno} \affiliation{\ROME}
\author {R.~Ent} \affiliation{\JLAB}
\author {F.~Garibaldi} \affiliation{\ROME}
\author {R.~Gilman} \affiliation{\RUTGERS} \affiliation{\JLAB}
\author {O.~Glamazdin} \affiliation{\KIPT}
\author {J.~Glister} \affiliation{\SMU} \affiliation{\DU}
\author {D.W.~Higinbotham} \affiliation{\JLAB}
\author {C.E.~Hyde-Wright} \affiliation{\ODU}
\author {Y.~Ilieva} \affiliation{\SCAROLINA}
\author {C.W.~de~Jager} \affiliation{\JLAB}
\author {X.~Jiang} \affiliation{\RUTGERS}
\author {M.K.~Jones} \affiliation{\JLAB}
\author {C.E.~Keppel} \affiliation{\HU} 
\author {E.~Khrosinkova} \affiliation{\KENT}
\author {E.~Kuchina} \affiliation{\RUTGERS}
\author {G.~Kumbartzki} \affiliation{\RUTGERS}
\author {B.~Lee} \affiliation{\SNU}
\author {R.~Lindgren} \affiliation{\UVA} 
\author {D.J.~Margaziotis} \affiliation{\CSU}
\author {D.~Meekins} \affiliation{\JLAB}
\author {R.~Michaels} \affiliation{\JLAB}
\author {K.~Park} \affiliation{\JLAB}
\author {L.~Pentchev} \affiliation{\WM}
\author {C.F.~Perdrisat} \affiliation{\WM} 
\author {E.~Piasetzky} \affiliation{\TELAVIV}
\author {V.A.~Punjabi} \affiliation{\NSU} 
\author {A.J.R.~Puckett} \affiliation{\MIT}
\author {X.~Qian} \affiliation{\DUKE}
\author {Y.~Qiang} \affiliation{\MIT}
\author {R.D.~Ransome} \affiliation{\RUTGERS}
\author {A.~Saha} \affiliation{\JLAB}
\author {A.J.~Sarty} \affiliation{\SMU}
\author {E.~Schulte} \affiliation{\RUTGERS}
\author {P.~Solvignon} \affiliation{\ARGONNE} 
\author {R.R.~Subedi} \affiliation{\KENT} 
\author {L.~Tang} \affiliation{\HU} 
\author {D.~Tedeschi} \affiliation{\SCAROLINA}
\author {V.~Tvaskis} \affiliation{\HU}
\author {J.M.~Udias} \affiliation{\UCM}
\author {P.E.~Ulmer} \affiliation{\ODU} 
\author {J.R.~Vignote} \affiliation{\IEM}
\author {F.R.~Wesselmann} \affiliation{\NSU}
\author {B.~Wojtsekhowski} \affiliation{\JLAB}
\author {X.~Zhan} \affiliation{\MIT}

\collaboration{The E03-104 Collaboration}
     \noaffiliation

\date{\today}

\begin{abstract}
We measured with unprecedented precision the induced polarization $P_{y}$ in 
$^4$He$(e,e^\prime\vec{p}\,)$$^3$H at $Q^{2} = 0.8$ (GeV$/c$)$^{2}$ and $1.3$ (GeV$/c$)$^{2}$. 
The induced polarization is indicative of reaction-mechanism effects beyond the 
impulse approximation. 
Our results are in agreement with a relativistic distorted-wave impulse approximation 
calculation but are over-estimated by a calculation with strong charge-exchange effects. Our 
data are used to constrain the strength of the spin independent charge-exchange term 
in the latter calculation. 
\end{abstract}
\maketitle

%%%%%%%%%%%%%%%%%%%%%%%%%%%%%%%%%%%%%%%%%%%%%%%%%%%%%%%%%%%%%%%%%%%%%%%%%
Why and to what extent the nucleon changes its structure while embedded in nuclear medium 
has been a long-standing question in nuclear physics, attracting experimental and 
theoretical attention. In this context, one of the hotly debated topics has been the 
interpretation of the quenching in the polarization-transfer 
double ratio, $(P_{x}^{'}/P_{z}^{'})_{^4He}/(P_{x}^{'}/P_{z}^{'})_{^1H}$ extracted from measurements 
of the polarization-transfer coefficients, $P_{x}^{'}$ and $P_{z}^{'}$, in elastic $\vec e p$ 
scattering and quasielastic scattering on ${^4}$He \cite{e93049,udias_all,rocco,e03104_trans}. 
In elastic $\vec e p$ scattering 
$P_{x}^{'}/P_{z}^{'}$ is directly proportional to the ratio of the 
electric and magnetic form factors of the proton, $G_{E}/G_{M}$ \cite{akhiezer}. 
In A$(\vec e,e^\prime\vec{p}\,)$B quasielastic scattering, the polarization-transfer 
ratio is expected to be sensitive to the 
form-factor ratio of the proton embedded in the nuclear medium. The polarization 
double ratio is then 
taken to emphasize differences between the in-medium and free values. For a ${^4}$He nucleus 
this double ratio was found to be quenched by 10\% \cite{e93049,e03104_trans}. This quenching 
could be due to conventional nuclear medium effects like nucleon off-shellness, 
meson-exchange currents (MEC), final-state interactions (FSI) but also to unconventional 
effects like modifications of the electric and magnetic form factors of the proton in the 
nuclear medium \cite{ff_med_mod}. However, an interpretation of a small quenching in 
the polarization-transfer double ratio as evidence of unconventional nuclear effects 
requires excellent control of conventional reaction mechanisms and hence remains, to 
some degree, model dependent. The induced polarization 
$P_{y}$, experimentally accessible along with $P_{x}^{'}$ and $P_{z}^{'}$, is 
a measure of conventional nuclear effects and offers vital constraints for the 
interpretation of the polarization-transfer double ratio.  

The polarization-transfer double ratio from Jefferson Lab experiments 
E93-049 \cite{e93049} and E03-104 \cite{e03104_trans} has been 
successfully modeled by two competing theoretical predictions: 
the relativistic distorted wave impulse approximation (RDWIA) calculation by the 
Madrid group \cite{udias_all} with medium-modified form factors from the quark-meson 
coupling (QMC) 
model \cite{ff_med_mod}, and the calculation of Schiavilla {\it et al.} \cite{rocco} 
which assumes free nucleon form factors but has different modeling of nuclear 
conventional effects, in particular the FSIs. Whether the two models 
give an accurate description of $P_{y}$ has become the key in the interpretation of 
the polarization-transfer double ratio. 
The large uncertainties of E93-049 $P_{y}$ measurements 
precluded any definite conclusion. Experiment E03-104 
provides the most precise measurements to date of the induced polarization 
in $^4$He$(e,e^\prime\vec{p}\,)$$^3$H and this letter presents the results. 

We report measurements of the induced polarization $P_{y}$ in the quasi-elastic 
reaction $^4$He$(e,e^\prime\vec{p}\,)\,^3$H, at four-momentum 
transfer, $Q^{2}$, of 0.8 and 1.3 (GeV$/c$)$^{2}$ and missing momentum, $p_{m}$, 
ranging from 0 to 160 MeV$/c$. 
A longitudinally polarized electron beam with flipping polarization direction 
and a current of 80 $\mu$A was incident on $^{4}$He and $^{1}$H targets and 
the scattered electron and recoil proton 
were detected in coincidence in two high-resolution spectrometer arms. 
The $^{4}$He target was chosen because its relative simplicity allows for 
realistic microscopic theoretical calculations while its high 
nuclear density increases the sensitivity to nuclear medium effects. 
The proton arm central momenta for the $^1$H$(\vec{e},e^\prime\vec{p}\,)$ reaction 
were adjusted in 2\% increments from -8\% to +8\% so that protons in elastic 
$\vec e p$ scattering had a similar 
coverage of the focal plane as in the $^4$He$(\vec{e},e^\prime\vec{p}\,)\,^3$H  
reaction \cite{e03104_trans}. 
These $\vec e p$ measurements provided a baseline for the comparison of in-medium 
to free proton polarizations and were also used to check for possible 
instrumental asymmetries. 

The polarized recoil protons traveled through the 
magnetic field of the spectrometer to the detector package used to measure 
the polarizations, the focal plane polarimeter (FPP) \cite{vina}. The spin 
precession of the protons was calculated using a well established model of the 
spectrometer's magnetic field \cite{e03104_trans}. In the FPP, the polarized protons 
scattered in a carbon block leading to azimuthal asymmetries. These 
asymmetries in combination with information on the proton 
spin precession and the carbon analyzing power were analyzed by means of a 
maximum likelihood method to obtain the induced polarization \cite{besset}.  

The extraction of the induced polarization $P_{y}$ is complicated by the presence 
of instrumental asymmetries. For the particular reaction that we studied, $P_{y}$ was 
expected to be small, $<$ 6\% \cite{e93049}. Thus even small instrumental asymmetries could 
constitute a significant background. The $^1$H data have been used to check 
for the presence of instrumental asymmetries. In the one-photon-exchange 
approximation $P_{y}$ 
in $^1$H$(e,e^\prime\vec{p}\,)$ is expected to be zero. The two-photon-exchange 
processes could yield a non-zero but rather small induced polarization, 
theoretical calculations 
predicting a value below 1\% \cite{wally,andrei} at our kinematics. 
When taking into account the analyzing power and the recoil proton spin transport, 
this will translate into an 
expectation for the physics azimuthal asymmetries of $<$ 0.4\% making any significant 
instrumental asymmetries easy to detect. 

We performed an extensive study to identify 
and correct for these asymmetries. The azimuthal distributions of the polarized 
protons are reconstructed from the track information provided by the FPP straw 
chambers located before (front) and after (rear) the carbon analyzer 
\cite{vina}. We engaged in a thorough check of the performance 
of the chambers and we found that inefficient regions and misalignments of the front 
and rear chambers lead to a contamination of the physics asymmetries. We devised 
a new tracking algorithm to allow track reconstruction even in inefficient regions 
and we developed a more precise alignment procedure to correct for 
misalignments. As a result, we see only small variations, at the sub-percent level, 
in the experimental azimuthal asymmetries for $^1$H$(e,e^\prime\vec{p}\,)$, 
with few outlyers up to 1\% in the few inefficient regions and at the 
edges of the FPP acceptance. On average, the experimental azimuthal asymmetries for 
$^1$H$(e,e^\prime\vec{p}\,)$  are at the sub-percent level.

To cancel out these residual instrumental asymmetries we obtain $P_{y}$ in               
$^4$He$(e,e^\prime\vec{p}\,)$$^3$H, $P_{y}$ in Table 1, 
as the difference of $P_{y}$ extracted from $^{4}$He, $P_{y}$(raw) in Table 1, 
and $^{1}$H data. Our systematic studies found the induced polarization thus extracted to be 
very robust on average and within the acceptance of the detector 
when binned in various kinematic variables. 
We reduced the systematic uncertainty on the $P_{y}$ extraction by a factor of four when 
compared to previous, similar measurements from E93-049 \cite{e93049}. Our data are used 
to put to stringent test state-of-the-art theoretical calculations and such comparisons 
are presented in what follows.

\begin{table*}[]
\centering
\caption[P_{y}]{The induced polarization $P_{y}$ from E03-104. 
$P_{y}$(raw) is the experimental value of the induced polarization in 
$^4$He$(e,e^\prime\vec{p}\,)$$^3$H. 
$P_{y}$ is the difference between the experimental values of $P_{y}$ in 
$^4$He$(e,e^\prime\vec{p}\,)$$^3$H and $^1$H$(e,e^\prime\vec{p}\,)$ which 
gives, in the absence of the induced polarization in $^{1}$H, the induced polarization in $^{4}$He. 
Stat. and syst. represent the statistical and systematic uncertainties, respectively.}
\begin{tabular}[c]{ccccccc}
\hline
\hline
$Q^2$         & $P_{y}$(raw) & stat. & $P_{y}$  & stat. & syst. \\
(GeV/$c$)$^2$ &                &       &           & &  \\
\hline
0.8 & $-$0.0366  & 0.0042 & $-$0.0415 & 0.0050 & 0.0050 \\
1.3 & $-$0.0394 & 0.0039  & $-$0.0373 & 0.0043 & 0.0058 \\
\hline
\hline
\end{tabular}
\label{tab:kin}
\end{table*}

\begin{figure}[htbp]
\vspace*{-0.2in}
\centering
\begin{tabular}{c}
\includegraphics[width=8.35cm]{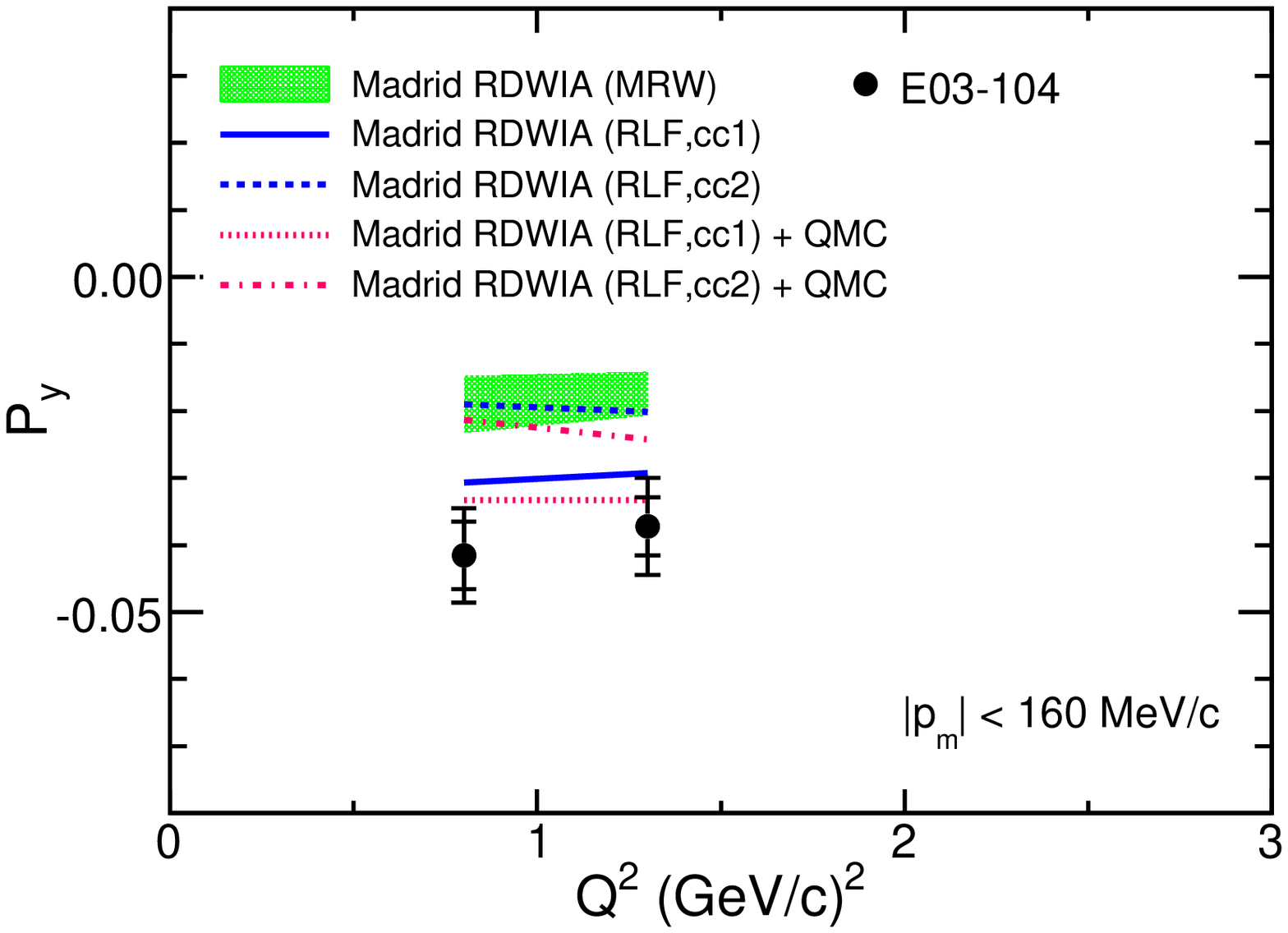} \\
\includegraphics[width=8.35cm]{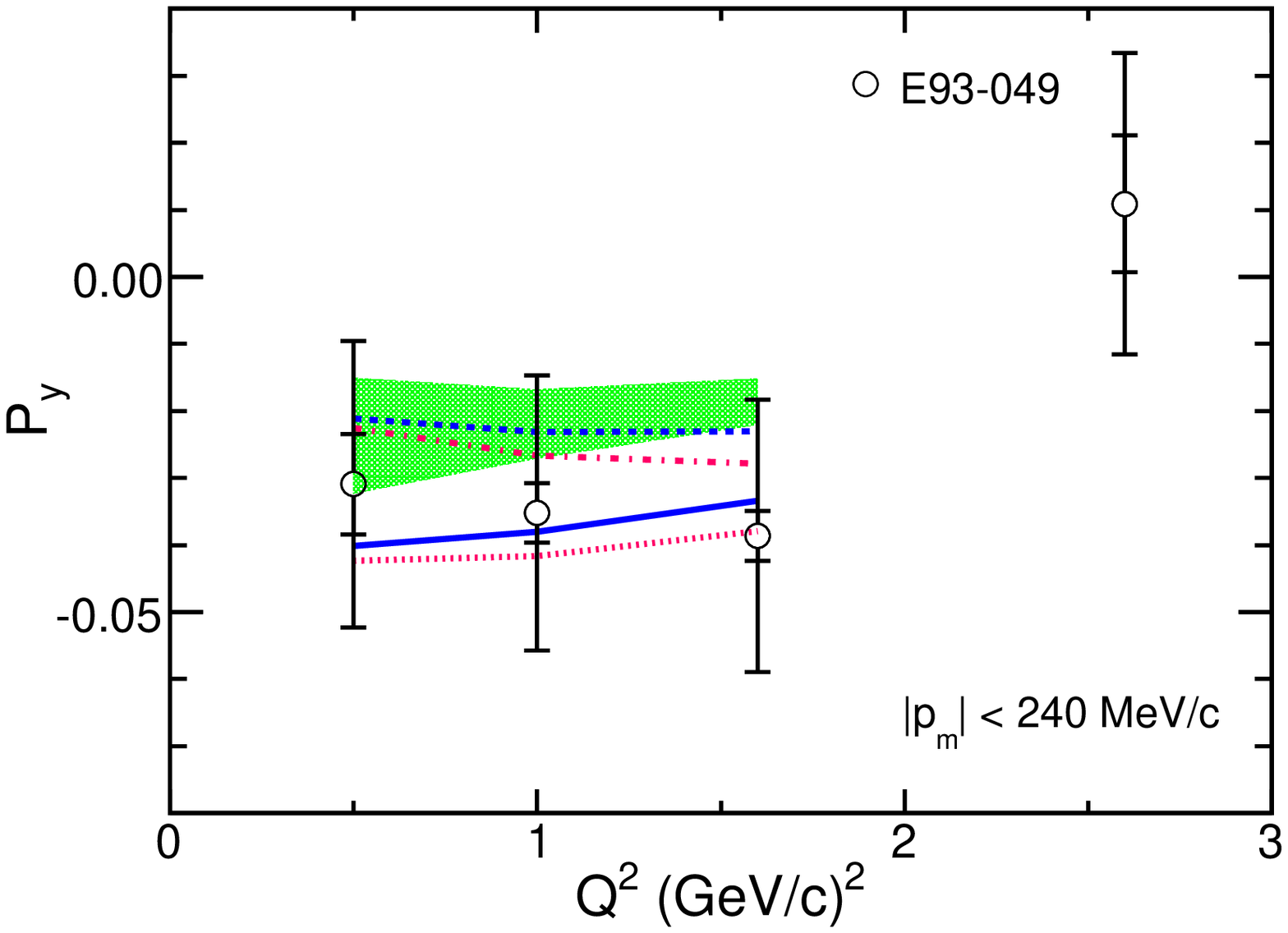} 
\end{tabular}
\linespread{0.5}
\caption[$p_{m}$ distributions.]{
{}(Color online) The acceptance averaged induced polarization $P_{y}$ in 
$^4$He$(e,e^\prime\vec{p}\,)$$^3$H as a function of $Q^{2}$ from 
E03-104 (upper panel) and E93-049 (lower panel) \cite{e93049}. The inner 
and outer error bars represent the statistical and total uncertainties, respectively. The 
green band displays the Madrid calculation \cite{udias_all} when using MRW as optical 
potential and {\it cc1} or 
{\it cc2} for the form of the nuclear current operator. For the solid and dashed curves RLF was used 
with {\it cc1} (solid) and {\it cc2} (dashed). The dashed-dotted and dotted 
curves were obtained from the dashed and solid ones, respectively, 
by including medium-modified form factors via the QMC model \cite{ff_med_mod}. 
Lines connect the acceptance-averaged theory calculations.}
\label{vertex}
\end{figure}

In Fig. 1 we show the induced polarization $P_{y}$ in $^4$He$(e,e^\prime\vec{p}\,)$$^3$H 
extracted from E03-104 (upper panel) together with earlier results 
from E93-049 \cite{e93049} (lower panel) and theoretical calculations from 
the Madrid group \cite{udias_all} (curves and band) which were averaged 
over the spectrometer acceptance. In the RDWIA, 
the nuclear current is calculated with relativistic wave functions for the initial bound and 
outgoing proton. The nuclear current operator can be of {\it cc1} or {\it cc2} 
forms \cite{deforest} depending on the prescription used to enforce current conservation. 
The final outgoing proton wave function is a solution 
of a Dirac equation with global optical potentials to account for FSIs. The optical 
potential models used are McNeil-Ray-Wallace (MRW) \cite{mrw} and Love-Franey (RLF) \cite{rlf}. 
In Fig. 1 the green band represents 
the Madrid calculation when MRW is used and the width of the band depends on the 
form of the nuclear current operator, {\it cc1} or {\it cc2}, with {\it cc1} giving a 
larger $P_{y}$ in absolute value. The blue solid and dashed curves represent the Madrid 
calculation when using the RLF optical potential and {\it cc1} (solid) or {\it cc2} (dashed).  
The older E93-049 experiment \cite{e93049} averaged over a larger range in missing momentum 
than E03-104, see Fig. 1, and $P_{y}$ is predicted to increase, in absolute value, with 
increasing missing momentum. This causes the 
apparent drop in the theory results at the kinematics of E93-049 when compared to those of 
E03-104.
The considerably reduced systematic uncertainties of the new results make possible 
a clear distinction between various theoretical prescriptions: the best description of the 
data is given by RDWIA (RLF,{\it cc1}).
The inclusion of medium-modified form factors via the QMC model \cite{ff_med_mod} slightly increases 
$P_{y}$, in absolute value, more so at large $Q^{2}$, as shown by the dashed-dotted 
and dotted curves. 

\begin{figure}[htbp]
\vspace*{-0.2in}
\centering
\begin{tabular}{c}
\includegraphics[width=8.5cm]{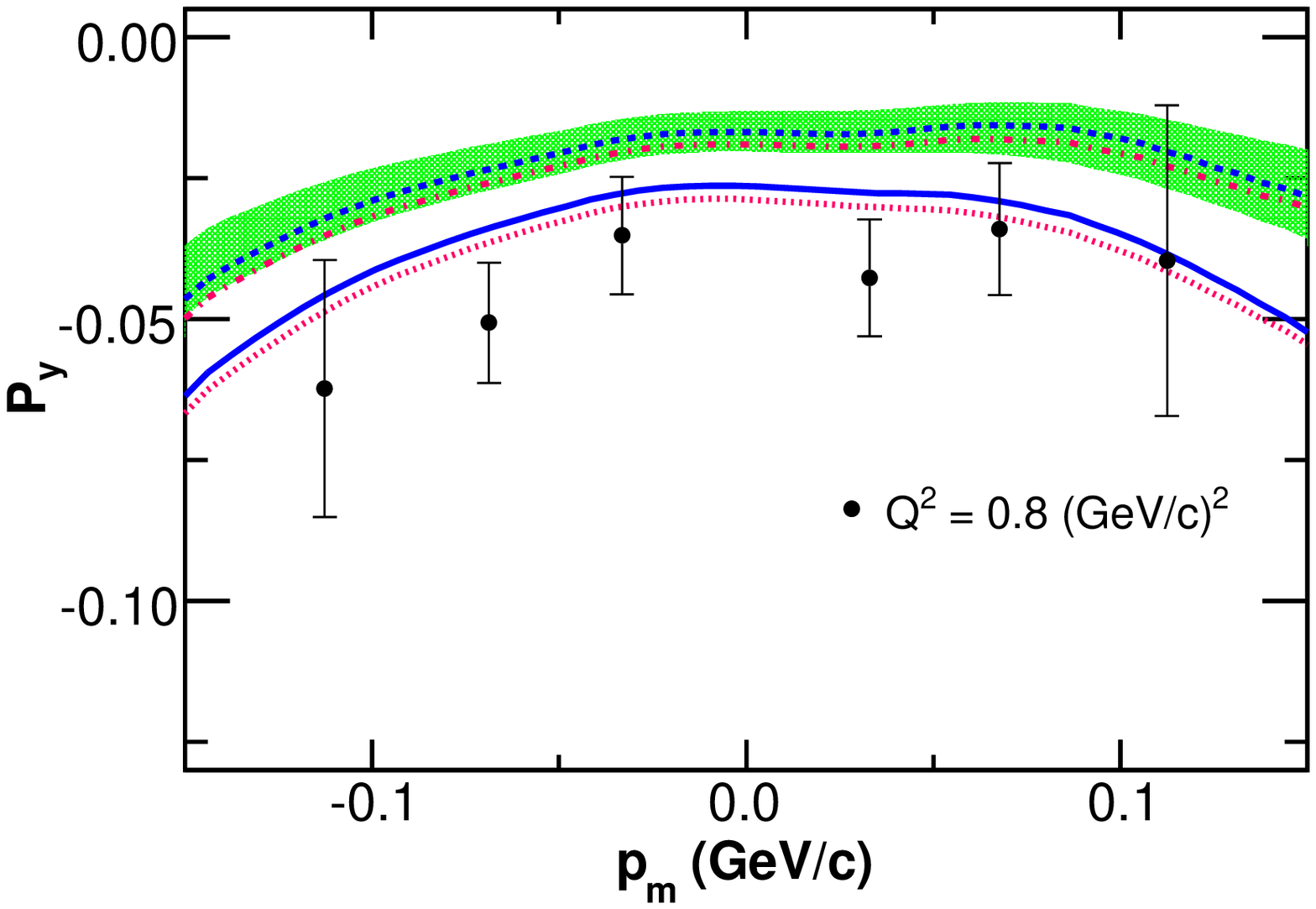} \\
\includegraphics[width=8.5cm]{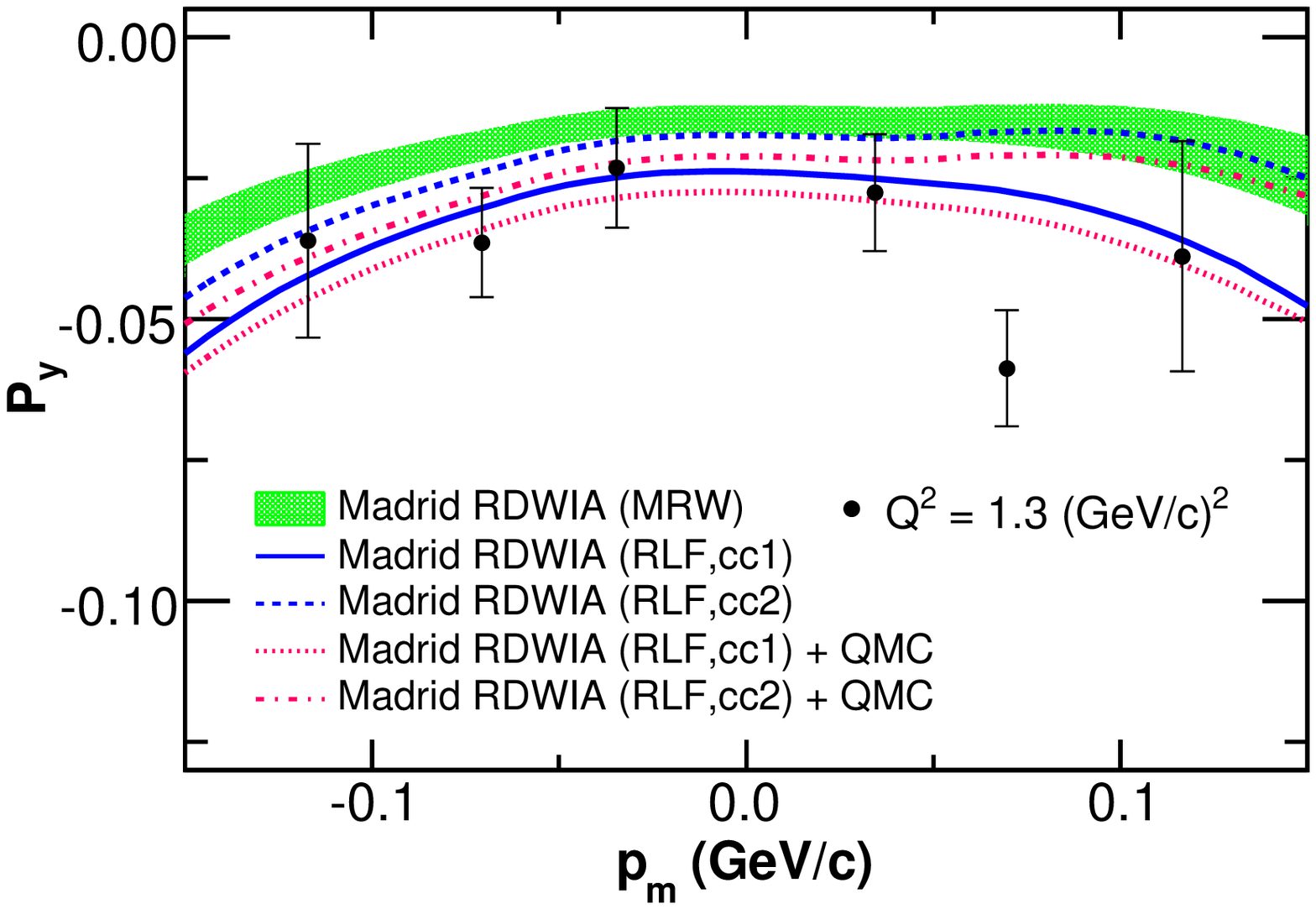} 
\end{tabular}
\linespread{0.5}
\caption[$p_{m}$ distributions.]{
{}(Color online) $P_{y}$ in $^4$He$(e,e^\prime\vec{p}\,)$$^3$H as a function of 
missing momentum $p_{m}$ from E03-104 (solid circles). Also shown are calculations from the 
Madrid group \cite{udias_all} (notations as in Fig. 1).}
\label{vertex}
\end{figure}

In Fig. 2 we present the distribution of the induced polarization $P_{y}$ as function of the 
missing momentum $p_{m}$. 
Our results are compared to the Madrid RDWIA calculation \cite{udias_all}. 
Overall, there is good agreement between data and the theoretical prediction. Both 
data and calculation show an increase in $P_{y}$ (in absolute value) with 
increasing $p_{m}$. The calculation predicts a stronger variation of $P_{y}$ 
in the range of $p_{m}$ from $-0.15$ GeV$/c$ to 0 GeV$/c$ compared to 0 GeV$/c$ 
to 0.15 GeV$/c$. Although there is some hint in the data that supports this behavior, 
especially at $Q^{2}$ = 0.8 (GeV$/c$)$^{2}$, the size of the statistical uncertainties 
preclude any definite conclusion. 

\begin{figure}
\includegraphics[width=8.5cm]{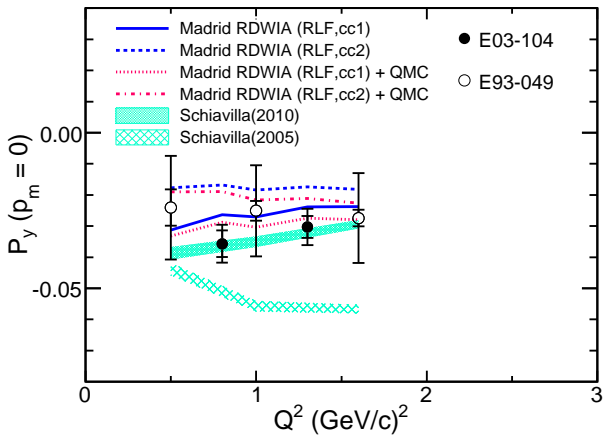}
\caption{(Color online) The induced polarization $P_{y}$ in $^4$He$(e,e^\prime\vec{p}\,)$$^3$H 
as a function of $Q^{2}$ extrapolated 
at missing momentum $p_{m}$ $\approx$ 0 from E03-104 (solid circles) and 
E93-049 (empty circles) \cite{e93049}. The inner and outer error bars represent 
the statistical and total uncertainties, respectively. Calculations from the Madrid group 
\cite{udias_all} (continuous and dashed curves) and from Schiavilla {\it et al.} 
\cite{rocco, rocco_2010} (bands) are also shown.}
\label{fig:F2}
\end{figure}
Another state-of-the-art theoretical calculation is the computation of Schiavilla {\it et al.} 
which uses variational wave functions for the bound three- and four-nucleon 
systems, non-relativistic MEC (2-body currents) and free nucleon form factors. The FSIs are 
treated within the optical potential framework and include both spin-independent and 
spin-dependent charge-exchange terms which play a crucial role in the prediction of $P_{y}$ 
\cite{rocco} for this calculation. The spin-independent charge-exchange term 
is constrained by $p$ + $^3$H $\to$ $n$ + $^3$He 
charge-exchange cross section data while the spin-dependent one is largely unconstrained \cite{rocco}.

In Fig. 3 our $P_{y}$ results are compared to the calculation of Schiavilla {\it et al.} 
\cite{rocco}. To facilitate this comparison our data have been corrected for 
the spectrometer acceptance as this theoretical calculation is only 
available at $p_{m} \approx 0$. This correction ($<$20\% for this experiment) was determined using 
the Madrid RDWIA (RLF,{\it cc1}) model 
because it offers a very good qualitative description of the $P_{y}$ dependence on $p_{m}$. 
The stability of the acceptance correction was studied by using other prescriptions within 
the Madrid RDWIA calculation and the variations were negligible when compared to the experimental 
uncertainties. The calculation from the Madrid group is also shown at $p_{m} \approx 0$. 
The remaining small but visible variation in the induced polarization between E93-049 and E03-104 
kinematics is due to the higher beam energies used in E93-049.
Although the prediction of Schiavilla {\it et al.}, Schiavilla(2005) in Fig. 3, offers a 
good description of the polarization-transfer double ratio \cite{e03104_trans}, 
it over-predicts, in absolute value, our measurements of $P_{y}$, especially at larger $Q^{2}$. 
This evident discrepancy prompted a revision of this calculation. The new calculation 
\cite{rocco_2010}, Schiavilla(2010) in Fig. 3, uses our $P_{y}$ results as additional constraints 
for the modeling of the charge-exchange terms. The calculation proved insensitive 
to variations of the spin-dependent charge-exchange term (especially at larger $Q^{2}$) 
and this remains largely unconstrained \cite{rocco_2010}. However, the spin-independent 
charge-exchange contribution has been modified to provide a good fit to our data. 
It remains to be verified whether the agreement with the charge-exchange 
cross section data from $p$ + $^3$H $\to$ $n$ + $^3$He is still maintained. 
This 2010 version of the calculation is also in good 
agreement with the polarization-transfer double ratio.    

The role of the charge exchange in $^4$He$(e,e^\prime\vec{p}\,)$$^3$H 
still needs to be clarified. $P_{y}$ in Schiavilla's 2010 
calculation \cite{rocco_2010} proves to be mostly sensitive to the charge-exchange 
spin-independent term, leaving the 
spin-dependent one still unconstrained; the Madrid group 
deems $P_{y}$ largely insensitive to both terms within the RDWIA frame work. On the other hand, the 
charge-exchange cross sections as predicted by the Madrid RDWIA calculation need 
to be gauged against data.
Possibly, a comparison of the induced polarization in $^4$He$(e,e^\prime\vec{p}\,)$$^3$H and 
$^4$He$(e,e^\prime\vec{n}\,)$$^3$He could cast some light on the role of charge exchange 
in this reaction.

To summarize, we measured with unprecedented precision the induced polarization $P_{y}$ in 
$^4$He$(e,e^\prime\vec{p}\,)$$^3$H at $Q^{2} = 0.8$ (GeV$/c$)$^{2}$ and $1.3$ (GeV$/c$)$^{2}$. For 
the first time the systematic uncertainties on this observable were reduced to a size 
comparable to the statistical uncertainties. 
We compared our results with theoretical calculations from the Madrid group \cite{udias_all} 
and Schiavilla {\it et al.} \cite{rocco,rocco_2010}. The Madrid RDWIA prediction describes 
well our data when RLF is used as optical potential and $cc1$ as form for the nuclear 
current operator. The 2005 prediction 
from Schiavilla {\it et al.} over-estimates our $P_{y}$ results (in absolute value) but 
gives a good description of the polarization-transfer double ratio \cite{e03104_trans}. 
Our induced polarization data have then been used to constrain the charge-exchange 
spin-independent term in the calculation and this 2010 version describes well both 
the induced polarization and the polarization-transfer double ratio. 
Our high-precision data point to the need to carefully consider both the charge-exchange 
spin-independent and spin-dependent terms in realistic calculations, to settle the extent 
of medium modifications in nucleon structure.

%%%%%%%%%%%%%%%%%%%%%%%%%%%%%%%%%%%%%%%%%%%%%%%%%%%%%%%%%%%%%%%%%%%%%%%%%

%%%%%%%%%%%%%%%%%%%%%%%%%%%%%%%%%%%%%%%%%%%%%%%%%%%%%%%%%%%%%%%%%%%%%%%%%
\begin{acknowledgments}

The collaboration wishes to acknowledge the Hall A technical staff and the Jefferson Lab 
Accelerator Division for their support. This work was supported by the U.S. Department of 
Energy and the U.S. National Science Foundation. Jefferson Science Associates operates 
the Thomas Jefferson National Accelerator Facility under DOE contract DE-AC05-06OR23177.

\end{acknowledgments}

%%%%%%%%%%%%%%%%%%%%%%%%%%%%%%%%%%%%%%%%%%%%%%%%%%%%%%%%%%%%%%%%%%%%%%%%%

\end{document}